\documentstyle[aps,amssymb,prl,multicol,times,graphicx]{revtex}
\draft
\begin{document}
\title{Spectral Shape of Relaxations in Silica Glass}
\author{Johannes~Wiedersich, Sergei V.~Adichtchev, Ernst~R\"{o}ssler}
\address{Physikalisches Institut, Universit\"{a}t Bayreuth, D-95440 Bayreuth, Germany}
\date{December 15, 1999}
\maketitle
   
\begin{abstract}
Precise low-frequency light scattering experiments on silica glass are presented, covering a broad 
temperature and frequency range 
($9\,\textrm{GHz}<\nu<2\,$THz). 
For the first time the spectral shape of 
relaxations is observed over more than one decade in frequency. The spectra show a power-law low-frequency 
wing of the relaxational part of the spectrum with an exponent $\alpha $ proportional to temperature in the 
range $30\,\textrm{K}<T<200\,$K.
A comparison of our results with those from acoustic attenuation
experiments performed at different frequencies shows that this
power-law behaviour rather well describes relaxations in silica over 9 orders of magnitude
in frequency.
These findings can be explained by a model of thermally activated 
transitions in double well potentials. 
\end{abstract}
\pacs{PACS: 78.35.+c,61.43.Fs,62.80.+f}

\begin{multicols}{2}
\narrowtext
Relaxations are a characteristic feature 
of glasses. They show up as a 
broad quasi-elastic contribution in neutron and light scattering spectra 
\cite{win:75,buc:88,sok:95}, 
and also as a damping of sound waves \cite{and:55,hun:74,hun:92a,kei:93,top:96} 
or as a dielectric loss \cite{fon:79}. 
Acoustic attenuation, or more precisely the internal friction $Q^{-1}$,
of silica glass (a-SiO$_2$) obtained at different
frequencies 
demonstrates the presence of a broad distribution of relaxation rates
\cite{and:55,hun:74}.
For temperatures above some 10\,K 
it is assumed that relaxations are due to thermally activated transitions in asymmetric double well potentials 
(ADWP's) \cite{hun:74,the:76,gpm:81},
where the same ADWP's relax via tunnelling at low temperatures ($T\lesssim1$\,K) \cite{var:72}.
Within this model it is possible to extract the distribution of barrier heights $g(V)$ from the 
experimental data \cite{hun:74,kei:93,top:96,fast1}. 

Although relaxations in silica have been extensively studied so far, 
a spectroscopic work that covers a broad frequency range is still missing. 
Acoustic and dielectric studies usually measure at a single frequency only, and
spectroscopic techniques like neutron and Raman scattering have typically
a low frequency limit around 100\,GHz and therefore cover only a small part of the 
relaxational contribution in a range where it is difficult to extract the 
spectral shape of relaxations. 
Recently, it has become possible to obtain low frequency
light scattering spectra of glasses at temperatures well below
$T_{\mathrm{g}}$, extending the spectral resolution to frequencies below
10\,GHz for some polymeric and anorganic glasses \cite{fast1,borox1}.

We refine this approach in order to study the much lower signal levels available for silica. 
Our data  
for the first time show the spectral shape of relaxations in silica
glass over a broad temperature and frequency range. This enables us to perform
a cross test on models about relaxations in glasses, regarding the frequency
and temperature dependence of relaxations. 
Furthermore we will compare our data with internal friction data to examine whether both 
techniques probe the same kind of relaxations. 

Depolarized inelastic light scattering spectra of a sample of Sup\-ra\-sil 300 (synthetic
silica, Heraeus, $<$1\,ppm of OH$^{-}$-groups) were obtained 
using an Ar$^{+}$\,laser (514.5\thinspace nm, 400\,mW) and a six-pass
Sandercock tandem Fabry-Perot interferometer \cite{lin:81}.
The sample was mounted in a 
dynamic Helium cryostat.
The Suprasil windows of the cryostat are
antireflection coated,
and the cold windows
are mounted tensionless in order to avoid tension induced birefringence.
Since the signal levels available from the scattering of silica are much lower
than for other glasses studied so far \cite{fast1,borox1},
the optical scheme was improved and optimized
for alignment on low intensities \cite{wie:diss}.
We use a 180$^{\circ}$ back-scattering geometry with a selection of depolarized
scattering.
Here the laser beam enters a Glan-Taylor polarizer (extinction 10$^{-6}$) as
the extraordinary beam. The polarized component of the light is
reflected and focused on the sample. The scattered light is collected by the
same lens and passes
through the same Glan-Taylor prism, transmitting the depolarized component.
The scattered light is collected within an angle of about 4$^{\circ}$.
Special care was taken to avoid
contributions to the signal from the windows of the cryostat and the polarizer.
This background has been carefully measured 
for all temperatures and free spectral ranges by recording spectra with and
without sample.
At all settings we find a background of about 3\% of the signal,
relating it to a contribution from the cold windows, which are 
close
to the focus of our lens, basically at the same temperature as the sample, and
of the same material as the sample.
Nevertheless, 
we carefully
eliminated its contribution 
by subtracting the signal from
accumulations without sample from all of our spectra.
In a double cross check, for some temperatures we also recorded spectra of
a sample of Heralux (fused silica, Heraeus) in a
near-to-back-scattering-geometry, avoiding any such contributions.
The thus recorded spectra 
agree with the results reported here within our 
accuracy.

We recorded spectra with the
free spectral ranges (FSR's) of
1000\,GHz and 150\,GHz over two spectral ranges on either side of the elastic
line. 
The experimentally determined finesse of the
spectrometer is better than 120 and typically about 140.
In order to suppress higher transmission
orders of the tandem (multiples of 20 FSR's could give contributions to the
signal for a Sandercock tandem FPI \cite{fast1,gap:99,cum:99,man:tandem}),
we use a prism in combination with either of two interference filters
of a width of 10\,nm and 1150\,GHz (FWHM).
The contrast of the spectrometer was determined 
to be better than $10^{9}$ at all frequencies. 

\begin{figure}[t]
\begin{center}
\includegraphics*[width=80mm]{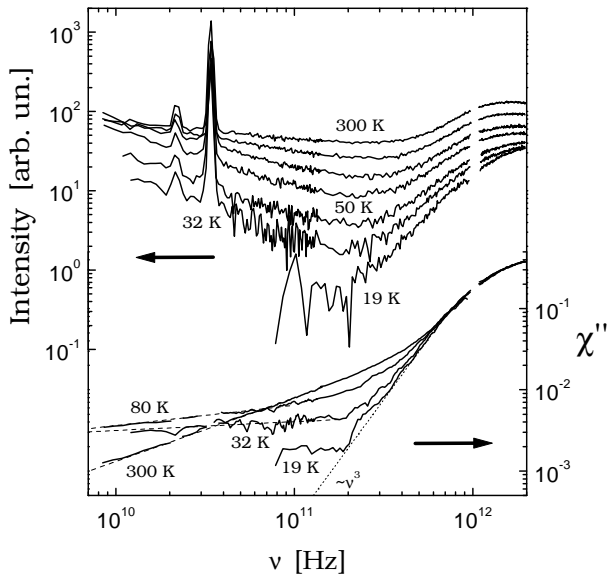}\vspace{1ex}
\caption[no]{Light scattering spectra of silica: Intensity spectra for the
temperatures 300\,K, 200\,K, 125\,K, 80\,K, 50\,K, 32\,K, and 19\,K
(from top to bottom, left scale) and the resulting susceptibility spectra
$\chi''(\nu)$ for the temperatures 300\,K, 80\,K, 32\,K, and 19\,K
(bottom, right scale, smoothing over 5 points) are shown. 
The dashed lines
correspond to a power-law fit of the low frequency part of the susceptibility
spectra. 
}
\label{spectrum}
\end{center}
\end{figure}

To further validate the absence of possible contributions from the
instrumental tail of the elastic line or from higher transmission orders of
the tandem, we measured spectra for both FSR's at a low temperature, 
$T=6$\,K. 
At this temperature the anti-Stokes part of the
spectrum should be almost zero because 
either the Bose factor or the signal itself are very low for this temperature. 
Indeed, the anti-Stokes part of the spectrum at this temperature shows no 
deviations from the dark count level of 2.47\,counts/s of our detector, 
demonstrating the absence of contributions of higher orders or from the
elastic line.
Any contributions from higher orders are especially problematic for the
smaller FSR, because the signal is much higher at frequencies of multiples of
20 of this FSR than in the range of interest: the signal increases with
increasing frequency, especially at low temperatures,
where the relaxational contribution is small (cf.\ Fig.~\ref{spectrum}).
For our 6\,K-spectrum in this region the sum of the signal and possible higher
order contributions is less than 0.3\,counts/s, ie, it is absent within the
precision of our experiment. Since the vibrational part, that could lead to
any parasitic contributions, rises much slower
with temperature than the signal from relaxations, it is clear that signals
from higher transmission orders do not disturb our spectra. 

The upper part of Figure~\ref{spectrum} displays the 
intensity of the 
depolarized back-scattering spectra of silica.
At high frequencies the Boson peak is observed. The signal passes through a
minimum and shows features of a central peak.
At frequencies of 35\,GHz and 20\,GHz
two narrow peaks are observed that correspond to the longitudinal and
transversal Brillouin lines and are due to a leakage from imperfect
polarisation and to the finite aperture, respectively.
Apart from the Brillouin lines, the central peak shows a power-law frequency
dependence, and its shape considerably changes with temperature.

\begin{figure}[t]
\begin{center}
\includegraphics*[height=52mm]{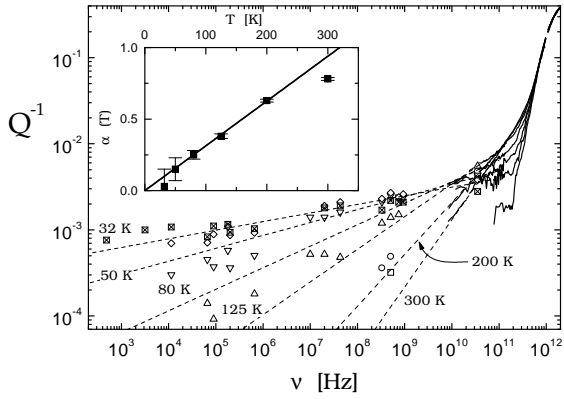}\vspace{1ex}
\caption{%
Light scattering susceptibility (solid lines) normalised to internal friction obtained 
from the Brillouin lines at 35\,GHz \protect\cite{vac:81}.
The dashed lines are power-law fits
to the light scattering data.
Symbols show the internal friction at the same temperatures
taken from the literature (references in the caption of
Fig.~\ref{rescale})
(empty squares: 300\,K, circles: 200\,K,
up triangles: 125\,K, down triangles: 80\,K, diamonds: 50\,K,
and crossed squares: 32\,K).
The inset shows the exponent $\alpha (T)$ of the low-frequency wing of the
light scattering susceptibility.
The solid line corresponds to a proportionality fit to
the data up to 200\,K and yields $\alpha = T / 319$\,K.
}
\label{alpha}
\end{center}
\end{figure}

In the susceptibility representation,  $\chi''(\nu)=I/(n(\nu )+1)$
of the Stokes side,
the trivial temperature dependence due to the Bose factor $n(\nu )$ is eliminated. 
The susceptibility data in Figure~\ref{spectrum} have been smoothed by adjacent
averaging over five points.
The presence of 
two contributions to the low-frequency light scattering in
glasses can be distinguished by the different temperature dependence:
the
vibrational contribution dominates at high frequencies and scales
with the Bose factor (ie, the susceptibility is temperature independent), 
wheras  the relaxational spectrum that dominates at
lower frequencies strongly changes with temperature.
As can be seen by the crossing of the spectra at different temperatures,
this is not simply due to an increase of the susceptibility with temperature
as would be expected in the case of
higher-order scattering processes. 
For all temperatures, the low-frequency wing of our spectra shows a power-law
behaviour.
The temperature dependence 
of the exponent,
$\alpha (T)$, is shown in the inset of Figure~\ref{alpha}:
$\alpha$ is proportional to temperature up to 200\,K, with $\alpha= T/319$\,K.

Let us see if this 
frequency and temperature dependence of relaxations can be 
described within our present understanding of glasses. 
In 1955 Anderson and B\"ommel attributed relaxations in silica to thermally
activated processes with a broad distribution of relaxation times
\cite{and:55}.
Theodorakopoulos and J\"ackle calculated the light scattering due to structurally relaxing two-state defects  
and show its relation to the acoustic attenuation \cite{the:76}. 
This approach was refined by Gilroy and Phillips \cite{gpm:81}, who consider
thermally activated transitions in ADWP's.
The potential wells are assumed to be the same that at lower temperatures
($T \lesssim 1$\,K) relax via tunnelling and are responsible for the low
temperature anomalies of glasses,
ie, the parameters for the potential wells are taken from the tunnelling model
\cite{var:72}.
Following the assumption that the distribution of asymmetry parameters $\Delta$ is flat,
the light scattering susceptibility $\chi''(\nu)$ and the
internal friction $Q^{-1}$ 
only depend on the distribution of barrier heights $g(V)$
\cite{the:76,gpm:81,kei:93,top:96}:
\begin{equation}
\label{intgV}
\chi\,''(\nu) \propto Q^{-1} \propto \int\limits_0^{\infty }
\frac{2\pi\nu\tau}{1+(2\pi\nu\tau)^2}\cdot g(V)\,dV.
\end{equation}
Here 
the relaxation time $\tau = \tau_0 \exp(V/k_{\mathrm{B}} T)$,where $\tau_0$ is the fastest
relaxation time that occurs.
Assuming an exponential distribution of barrier heights,
$g(V)=V_0^{-1}\exp (-V/V_0)$, the model predicts for the low frequency wing of
the relaxation spectrum, $\nu \ll (2\pi \tau_0)^{-1}$, a power-law
susceptibility spectrum with an exponent
proportional to temperature \cite{gpm:81}:
$\alpha = k_{\mathrm{B}} T/V_0$. 
The inset of
Figure~\ref{alpha} shows that the exponents of our silica data agree with the
model for temperatures below 200\,K and we obtain $V_0/k_{\mathrm{B}}=319$\,K.
At 300\,K the observed exponent is less than expected from the model. 
Since the exponent of the susceptibility 
cannot become
higher than 1 for 
\noindent
relaxation processes and is already close to 1 at room
temperature, the deviation at 300\,K comes as no surprise.
Indeed at higher temperatures up to $T_{\mathrm{g}}$ it is found that the
exponent remains constant at 1 \cite{sio2HT}.
The simple assumption of thermally activated relaxations over barriers with an
exponential distribution of barrier heights can therefore well account for the
observed temperature dependence of the spectral shape of the low-frequency
light scattering data of silica at temperatures up to 200\,K.

Within the model of Theodorakopoulos and J\"ackle the distribution of activation energies is 
reflected in the light scattering susceptibility just in the same manner as in the internal friction
(ie, $\chi''(\nu)\propto Q^{-1}$, cf.\ Eq.~\ref{intgV}) \cite{the:76,gpm:81}.
Taking the internal friction at a frequency that is also accessible to
light scattering spectroscopy,
the proportionality constant can be eliminated in order to directly compare
the results of both techniques.
In Figure~\ref{alpha} we plot the susceptibility data of Figure~\ref{spectrum}
scaled to the internal friction measured at 35\,GHz \cite{vac:81};
a single factor can be used for all spectra, showing that the temperature
dependence of $\chi''(\nu)$
around 35\,GHz is the same as that of the internal friction.
Acoustic data from different experiments
have a scatter of about a factor of two, and within
a factor of two follow the extrapolations of the light scattering data.
It appears that relaxations indeed show up in the same manner for both techniques, and that 
with good approximation
the spectral shape of
relaxations in silica shows a power-law behaviour with an exponent
$\alpha = T/319\,$K over the whole available frequency range
$500\,\textrm{Hz}<\nu<500$\,GHz.

This can be further demonstrated by comparing the distributions $g(V)$ of the barrier heights obtained from the different experiments. 
Equation~\ref{intgV} can be further simplified, assuming 
a broad
distribution of barrier heights $g(V)$. In this case the
\begin{figure}[t]
\begin{center}
\includegraphics*[height=52mm]{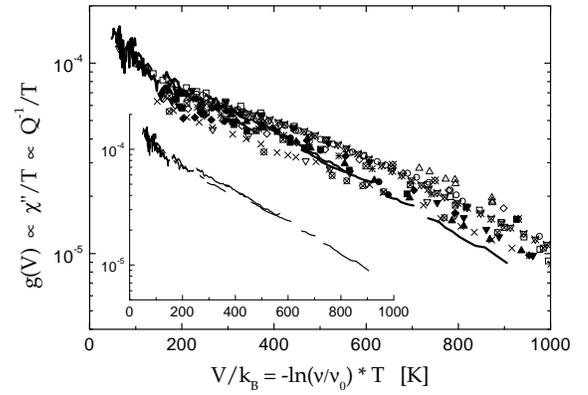}\vspace{1ex}%
\caption{%
Distribution of barrier heights $g(V)$ as obtained from the different techniques at
different temperatures and frequencies by rescaling the data. 
Solid lines: light scattering; symbols: internal friction 
(solid squares: 35\,GHz \protect\cite{vac:81},
solid circles: 930\,MHz \protect\cite{jon:64},
solid up triangles: 748\,MHz \protect\cite{jon:64},
solid down triangles: 507\,MHz \protect\cite{jon:64},
solid diamonds: 330\,MHz \protect\cite{jon:64},
open squares: 43\,MHz \protect\cite{bar:82},
open circles: 20\,MHz \protect\cite{and:55},
open up triangles: 10\,MHz \protect\cite{fin:54},
crosses: 660\,kHz \protect\cite{gpm:81},
open down triangles: 201\,kHz \protect\cite{fin:54}, 
open diamonds: 180\,kHz \protect\cite{kei:93}, 
crossed squares: 90\,kHz \protect\cite{cah:89},
crossed circles: 66\,kHz \protect\cite{fin:54},
crossed down triangles: 11.4\,kHz \protect\cite{hun:92a},
crossed circles: 3170\,Hz \protect\cite{ray:84},
stars: 484\,Hz \protect\cite{ray:84}).
The inset shows the rescaling of the light scattering data alone 
($T=200$\,K, 125\,K, 80\,K, 50\,K, and 32\,K). 
}
\label{rescale}
\end{center}
\end{figure}
\noindent
susceptibility
spectrum $\chi''(\nu)$ and the internal friction $Q^{-1}(T)$ for
$2\pi\nu\tau_0 \ll 1$ directly reflect the distribution of correlation
times, and thus the distribution $g(V)$ \cite{top:96}:
\begin{equation}
 \chi'' \propto Q^{-1} \propto T\,g(V), \,
 \textrm{where}\,\,
 V = k_{\mathrm{B}}T\, \ln(1/2\pi\nu\tau_0).
\end{equation}
Assuming that $g(V)$ is temperature independent and that
thermally activated transitions determine the light scattering spectra
and the internal friction at temperatures above some 10\,K, this
distribution of barrier heights can be directly extracted from the
data by rescaling the axes with $T$. Explicitly, we multiply the
$\log(\nu)$-axis by $T$ and divide the $Q^{-1}$ and $\chi''$ axes by
$T$. 

Then a master curve for $g(V)$ should result for the rescaled 
$\chi''(\nu)$ spectra and of the acoustic data
$Q^{-1}(T)$ for the different frequencies.
Since the amplitude of the light scattering data has already been fixed with
respect to the acoustic data,
the only parameter for the rescaling is
$\nu_0=(2\pi\tau_0)^{-1}$. Unlike the acoustic experiments
our spectra cover this cut-off frequency
at the transition from the power-law relaxational contribution to the vibrations in the range of the boson peak (cf.\ Fig.~\ref{spectrum}).
From the light scattering
spectra we obtain $\nu_0 = 800$\,GHz 
(ie, $\tau_0=0.2$\,ps), which is in good agreement with previous results taken from acoustic data 
\cite{hun:74,the:76}, 
and use this value for
the rescaling of
all the data.
For the rescaling of the light scattering data we use the frequency range where the power-law holds and where the 
contribution from vibrations to the spectra is negligible. 
In principle, the distribution $g(V)$ can be obtained from a single experiment at one frequency 
(cf.\ eg, \cite{kei:93,top:96}).
The light scattering experiment, however, probes the temperature and frequency dependence of the susceptibility 
simultaneously, allowing a cross test of the model \cite{fast1}.

Figure~\ref{rescale} shows the result of this rescaling procedure.
The light scattering data form the different temperatures show an excellent agreement and the exponential distribution as discussed above. 
Within the scatter of the acoustic data of about a factor of two, we
obtain a master curve for all the available data ($10\,\textrm{K}<T<200$\,K).
The rescaling of the 
abscissa involves both the temperature and
the frequency of the data. Therefore at a fixed barrier height of say
$V/k_{\mathrm{B}}=200$\,K, we compare data that have been obtained at a
temperature of 200\,K and a frequency of about 200\,GHz with those
obtained at a temperature of 10\,K and a frequency of 500\,Hz. 
Considering this range, it is indeed surprising that the scaling works so well.
The rescaled acoustic data cover the range both above and below the temperature of the loss peak found in 
silica at temperatures between some 25 and 130\,K \cite{hun:74,hun:92a}.
(In Fig.~\ref{alpha} the precence of this peak is not seen at low frequencies, 
because the lowest temperature plotted is 32\,K.)
Taking acoustic data alone, it was found that a modified Gaussian distribution
describes the data better than the exponential distribution obtained
from our light scattering data \cite{kei:93}. 
This is demonstrated by
the fact that some of the acoustic data show a slight curvature on the semi-logarithmic plot in 
Figure~\ref{rescale}, which is absent for the rescaled light scattering data \cite{note}. 
However, the simple model with only two parameters, $\tau_0$ and
$V_0$, rather well describes
relaxations in silica over a broad temperature and
frequency range.

In conclusion, we extend existing light scattering data on silica by
about one order of magnitude to lower frequencies, covering a broad
range in temperature down to some 20\,K. Our data for the first time
reveal the spectral shape of relaxations over more than one decade in
frequency, where relaxations clearly dominate over the vibrational
contribution to the spectrum.
The relaxations show a power-law spectral shape at low frequencies with
an exponent $\alpha$ proportional to temperature.
For the temperature range below 200\,K our data are in good quantitative
agreement with a model attributing
relaxations in glasses to thermally activated transitions with an 
exponential distribution of barrier heights $g(V)$.
We further compare the light
scattering susceptibility $\chi''(\nu)$ with the internal friction 
$Q^{-1}$ and find that 
within a factor of two 
the 
power-law behaviour
extends down to frequencies of some 500\,Hz.
Therefore relaxations in silica within the temperature range
$10\,\textrm{K}\lesssim T \lesssim 200\,\textrm{K}$ can be described by two single
parameters, 
$\nu_0=800$\,GHz and $V_0/k_{\mathrm{B}}=319$\,K, 
for all frequencies
extending up to the onset of the boson peak.
Although it is known that this model in its simplest form presented
here does not work for all
glasses \cite{fast1,borox1},
we believe that our demonstration
that the model 
works remarkably well for silica,
a paradigmatic glass, 
should be considered in any refinement of models describing relaxations
in glasses.

We thank V.\,N. Novikov and N.\,V. Surovtsev for many illuminating discussions and the 
DFG (SFB 279) for financial support. 
J.\,W.\ appreciates financial support from H. Wiedersich. 

\bibliographystyle{unsrt}

\begin{thebibliography}{20}

\bibitem{win:75}
G.~Winterling.
\newblock {\em Phys. Rev. B}, {\bf{12}}:2432, 1975.

\bibitem{buc:88}
U.~Buchenau, H.~M. Zhou, N.~Nucker, K.~S. Gilroy, and W.~A. Phillips.
\newblock {\em Phys. Rev. Lett.}, {\bf{60}}:1318, 1988.

\bibitem{sok:95}
A.~P. Sokolov, U.~Buchenau, W.~Steffen, B.~Frick, and A.~Wischnewski.
\newblock {\em Phys. Rev. B}, {\bf{52}}:R9815, 1995.

\bibitem{and:55}
O.~L. Anderson and H.~E. B\"ommel.
\newblock {\em J. Am. Ceram. Soc.}, {\bf{38}}:125, 1955.

\bibitem{hun:74}
S.~Hunklinger.
\newblock {\em Ultrasonics Symposium Proceedings}, IEEE New York, pp. 493--501, 1974;
S.~Hunklinger and M.~v.~Schickfus in 
{\em Amorphous Solids: Low-Temperature Properties}, 
edited by W.\,A.~Phillips,  
pages 81--106.
Springer-Verlag, Berlin, 1981.

\bibitem{hun:92a}
D.~Tielb\"urger, R.~Merz, R.~Ehrenfels, and S.~Hunklinger.
\newblock {\em Phys. Rev. B}, {\bf{45}}:2750, 1992.

\bibitem{kei:93}
R.~Keil, G.~Kasper, and S.~Hunklinger.
\newblock {\em J. Non-Cryst. Sol.}, {\bf{164-166}}:1183, 1993.

\bibitem{top:96}
K.~A. Topp and D.~G. Cahill.
\newblock {\em Z. Phys. B}, {\bf{101}}:235, 1996.

\bibitem{fon:79}
J.~Fontella, R.~L. Johnston, G.~H. Sigel, and C.~Andeen.
\newblock {\em J. non-cryst. Sol.}, {\bf{31}}:401, 1979.

\bibitem{the:76}
N.~Theodorakopoulos and J.~J\"ackle.
\newblock {\em Phys. Rev. B}, {\bf{14}}:2637, 1976;
J.~J\"ackle in
{\em Amorphous Solids: Low-Temperature Properties}, 
edited by W.\,A.~Phillips,  
pages 135--160.
Springer-Verlag, Berlin, 1981.

\bibitem{gpm:81}
K.~S. Gilroy and W.~A. Phillips.
\newblock {\em Phil. Mag. B}, {\bf{43}}:735, 1981.

\bibitem{var:72}
P.~W. Anderson, B.~I. Halperin, and C.~M. Varma.
\newblock {\em Phil. Mag.}, {\bf{25}}:1, 1972;
W.~A. Phillips.
\newblock {\em J. Low Temp. Phys.}, {\bf{7}}:351, 1972.

\bibitem{fast1}
N.~V. Surovtsev, J.~A.~H. Wiedersich, V.~N. Novikov, E.~R\"ossler, and A.~P.
  Sokolov.
\newblock {\em Phys. Rev. B}, {\bf{58}}:14888, 1998;

\bibitem{borox1}
N.~V. Surovtsev, J.~Wiedersich, E.~Duval, V.~N. Novikov, E.~R\"ossler, and
  A.~P. Sokolov.
\newblock {\em J. Chem. Phys.}, accepted.

\bibitem{lin:81}
S.~M. Lindsay, M.~W. Anderson, and J.~R. Sandercock.
\newblock {\em Rev. Sci. Instrum.}, {\bf{52}}:1478, 1981.

\bibitem{wie:diss}
J.~Wiedersich.
\newblock PhD thesis, Universit\"at Bayreuth.
\newblock in preparation.

\bibitem{gap:99}
J.~Gapi\'nski, W.~Steffen, A.~Patkowski, A.~P. Sokolov, A.~Kisliuk,
  U.~Buchenau, M.~Russina, F.~Mezei, and H.~Schober.
\newblock {\em J. Chem. Phys.}, {\bf{110}}:2312, 1999.

\bibitem{cum:99}
H.~C. Barshilia, G.~Li, G.~Q. Shen, and H.~Z. Cummins.
\newblock {\em Phys. Rev. E}, {\bf{59}}:5625, 1999.

\bibitem{man:tandem}
J.~R. Sandercock.
\newblock {\em Operator Manual for Tandem Interferometer}, 1993.

\bibitem{sio2HT}
J.~Wiedersich, N.~V. Surovtsev, and N.~Bagdassarov.
\newblock unpublished results.

\bibitem{vac:81}
R.~Vacher, J.~Pelous, F.~Plicque, and A.~Zarembowitch.
\newblock {\em J. Non-Cryst. Solids}, {\bf{45}}:397, 1981.

\bibitem{jon:64}
C.\,K.~Jones, P.\,G.~Klemens, and J. A. Rayne.
\newblock {\em Phys. Lett.}, {\bf{8}}:31, 1964.

\bibitem{bar:82}
U.~Bartell and S.~Hunklinger.
\newblock {\em J. Phys. Colloq}, {\bf{43}}:C9--489, 1982.

\bibitem{fin:54}
M.~E. Fine, H.~van Duyne, and Nancy~T. Kenney.
\newblock {\em J. Appl. Phys.}, {\bf{25}}:402, 1954.

\bibitem{cah:89}
D.~G. Cahill and J.~E.~Van Cleve.
\newblock {\em Rev. Sci. Instrum.}, {\bf{60}}:2706, 1989.

\bibitem{ray:84}
A.~K. Raychaudhuri and S.~Hunklinger.
\newblock {\em Z. Phys. B}, {\bf{57}}:113, 1984.

\bibitem{note}
These slight 
differences cannot be resolved by an extended analysis, 
evaluating Eq.~\ref{intgV} numerically. 
They will be discussed in more detail in an extended publication. 
\end{thebibliography}

\end{multicols}
\end{document}